# AI Approaches in Processing and Using Data in Personalized Medicine

Mirjana Ivanovic[1][0000-0003-1946-0384] and Serge Autexier[2][0000-0002-0769-0732],

Miltiadis Kokkonidis [3]

[1] University of Novi Sad, Faculty of Sciences, Novi Sad, Serbia
[2] German Research Center for Artificial Intelligence (DFKI), Bremen Site, Germany
[3] NETCOMPANY-INTRASOFT S.A., Luxembourg, Luxembourg
`mira@dmi.uns.ac.rs, serge.autexier@dfki.de,`
`Miltiadis.KOKKONIDIS@netcompany-intrasoft.com`

**Abstract.** In modern dynamic constantly developing society, more and more people suffer from chronic and serious diseases and doctors and patients need special and sophisticated medical and health support. Accordingly, prominent health stakeholders have recognized the importance of development of such services to make patients' life easier. Such support requires the collection of: huge amount of patients' complex data (clinical, environmental, nutritional, daily activities…), variety of data from smart wearable devices, data from clothing equipped with sensors etc. Holistic patient's data must be properly aggregated, processed, analyzed, and presented to the doctors/caregivers to recommend adequate treatment and actions to improve patient's health related parameters and general wellbeing.
Advanced artificial intelligence techniques offer the opportunity to analyze such big data, consume them, and derive new knowledge to support (personalized) medical decisions. New approaches like those based on advanced machine/deep learning, federated learning, transfer learning, explainable artificial intelligence open new paths for more quality use of health and medical data in future. In this paper, we will present some crucial aspects and characteristic examples in the area of application of a range of artificial intelligence approaches in (personalized) medical decisions.

## 1 Introduction

We are witnesses of more and more sick population and it is necessary to take care of development of sophisticated multi-disciplinary approaches for medical/clinical and healthcare diagnoses and treatments [7], [21]. Consequently, development of helpful medical services is getting crucial traction in medical innovation, at both research institutions and companies. The importance of improvements of patients' health related quality of life (QoL) parameters are also widely recognized in therapies and follow-ups of serious diseases survivors.

Cancer patients experience a serious disruption of QoL parameters (fatigue, pain, psychological difficulties, appetite loss, sexual problems and so on). Additionally, they experience also "usual" problems like the majority of the population (anxiety, stress, sleep disorders, psychological problems, mental impairment and so on) during active oncological treatment period.

To support the development of sophisticated software services that can help patients to successfully cope with everyday activities it is necessary to find ways to collect and properly integrate wide spectra of complex patient data (apart from traditional clinical data also data collected from smart wearable devices, environmental and wearable sensors, nutritional data, and so on). Health-related data should be aggregated in such forms that obtain adequate, useful, and reliable conclusions after processing. Results achieved after data processing should be presented to the doctors/clinicians in understandable and friendly form [7].

Modern, emergent approaches in collecting, processing, and analyzing patient's data support more appropriate interventions, and usually more tailored and personalized treatments [4], [10]. In this paper we will present the current state-of the-art in developing medical and clinical platforms and frameworks and discuss crucial aspects, functionalities, and characteristic examples in the area of application of a range of artificial intelligence approaches in (personalized) medical decisions [23], [24].

The rest of the paper is organized as follows. In Section 2, different sources of patients' medical and health-related big data are briefly discussed. Section 3 considers some emergent artificial intelligence approaches that support quality medical decisions. Two characteristic medical decision support systems are presented. Concluding remarks and future trends in processing big medical data are pointed out.

## 2  Different Sources of Patients' Medical Big Data - Collection and Processing

Electronic Health Record (EHR) is usually seen as basic source of information for any patient. It keeps data of several important aspects of a patient (like clinical information, diagnoses, medication, …). For more reliable follow-ups of patient's health, it is necessary to consider also other data sources and in modern medical data processing they also can include so called Patient Health Record (PHR). A PHR usually contains the same (or similar) kinds of information as an EHR but it is managed by patient.

For example, the CrowdHEALTH project [5] is oriented to the combination of patient's data from various sources to benefit from community knowledge and form Holistic Health Record (HHR). As a result of this project an integrated holistic platform is developed and it incorporates big data management mechanisms to support the logical pipeline of data managemen [13].

After identifying and considering different patient's data sources the next step in medical systems/platforms/frameworks is to find adequate ways and techniques to better acquire, manage, model, and process this data in order to achieve as much as possible, high-quality outcomes and results. Such result, based on huge amounts of data, should be exploited and presented in a user friendly way to doctors/clinicians, caregivers, or even to patients.

Another interesting approach oriented towards use of complex patient's data and processing it by application of modern AI/ML approaches is developing under BD4QoL project [2]. The focus in the project is on implementation of a personalized management of head and neck cancer survivorship by providing doctors and survivors with an unobtrusive, privacy compliant, real-time monitoring.

Measuring cancer patient's QoL is about understanding the impact of cancer and how well people are living after their diagnosis and treatment. This includes a wide range of concerns, such as people's emotional or social wellbeing, finances, and ongoing physical problems, such as tiredness, sleep disorders, and pain.

We can conclude that there is trend in medical decision support systems to integrate "traditional" medical and health data sources with novel ones that include data from smart and wearable devices, IoT and sensors generated data, open data, environmental data, etc.).

The rapid development of information communication technologies, applications of IoT and pervasive smart environments in our everyday life promotes the frequent use of different smart wearable devices for monitoring and measuring some health parameters [8]. Constant improvements and development of such devices impose that they should satisfy specific requirements.

If we concentrate on cancer patients, then it is evident that in several last decades the number of cancer survivors are increasing. Thus, there is a need to develop medical systems with tailored, personalized services that will help in improving patients' QoL parameters. So, it is important to include in patient's medical records their personal experiences. So, in contemporary medical decision support systems a number of questionnaires/tools to measure cancer patient's individual views of his/her health status should be considered.

Modern societies are getting more and more "smart". Especially, wearable devices play an important primary role to establish and maintain a connection between patients and doctors which offers a great potential to support the quality of medical treatment and recommendations. Additionally acquired complex data generated in smart environments and with wearable devices offers numerous opportunities in mobile health applications [18] or the development of complex IoT sensing-based health monitoring systems [6] [14].

## 3 Emergent Artificial Intelligence Approaches for Supporting Quality Medical Decisions

With appropriate medical treatment and support more and more people suffering from different critical diseases are living and normally go about their everyday routine activities. Also, more than ever people are living with and beyond cancer. Receiving adequate treatment tailored to their needs patients can keep and even increase their positive experience and QoL.

Personalized medical services for patients with similar needs are adequate therapies, decisions, interventions, and recommendations adjusted to their specific health status [1], [7]. Therapeutic strategy for "the right person at the right time" can support improvement and efficiency of the treatment, reduce possible side effects and increase the QoL.

QoL parameters are getting essential for cancer survivors. They influence the development of adequate services for person-centered monitoring and follow-up planning. Complex patient's data collected from multiple sources should be processed to be used in improving personalized treatment. Powerful AI/ML approaches are essential instruments for quality data processing that lead to better predictions, interventions and good health status. However, before applying AI/ML techniques diverse data must be prepared in an adequate way (i.e., aggregated, processed, analyzed). Contemporary AI approaches: (deep) ML [7], explainable AI (XAI), image processing (IP), natural language processing (NLP), agent technologies [10], robots, and so on, immensely influence the development of medical systems/platforms/frameworks. Such holistic systems should: improve post-treatment patients' health status and QoL; follow-up the patients to meet their needs and make their everyday life bearable; but also help in predicting the status of new patients.

Many large projects focus on cancer patients and better QoL parameters based on their available complex data. Considering some of them (e.g. https://oncorelief.eu/, https://www.gatekeeper-project.eu/, http://www.bd2decide.eu/, https://ascape-project.eu/, etc.), we outline a "typical Health AI system". Such a system is composed of several subsystems.

**1) Data Management subsystem** that is responsible for secured patients data collection from multiple sources usually taking care of anonymization [22]. They are also focused on the aggregation of heterogeneous data, their transformation in some of widely used clinical data standards which address different aspects [21] but also to prepare them in formats appropriate for AI/ML processing. The essential tasks of this subsystem are focused on multiple sources data collection and its preparation for AI/ML processing. During these activities privacy preservation of patients' data must be guaranteed.

**2) AI/ML subsystem** is tightly related and connected to the Data Management subsystem. This subsystem, for which suitable AI/ML methods are selected to be applied, is responsible for comprehensive data processing and analyses of computed results. After data processing important and influential features/parameters are discovered, characteristic patterns of behaviors are noticed, powerful predictive models are generated. Predictive models produce quality predictions, interventions, treatment recommendations that should be presented to doctors/caregivers/patients.

This subsystem usually includes Big Data Analytics and Modelling and it is the central in a medical system and represents the logical link between the data management part and interface part. The AI/ML subsystem uses a variety of ML algorithms based on available modern ML frameworks.

**3) Intelligent/Smart Interface subsystem** usually is the connection between patient's data and results achieved by AI/ML subsystem. Depending on the main aim of a medical system this subsystem can generate different forms of interfaces for doctors, caregivers, but also for patients. Interface uses generated AI/ML models and suggests adequate, personalised treatments, interventions, various actions, activities, nutrition and so on.

In typical medical systems various predictive models are generated to support personalized medical decisions. To make results of predictive models more understandable to doctors and other end-users recently XAI methods are using [9], and different ways of data visualization are applied.

Depending on general organization and use of specific medical systems/platform/frameworks, it can incorporate and consider patients' datasets from arbitrary number of hospitals, train and lately use AI/ML predictive models considering common knowledge gained from all available datasets. Such types of systems adopt federate style of data processing, models training and using achieved powerful AI/ML predictive models. In such an approach sensitive patient's data remains decentralized and FL keeps the data at its local edge nodes (hospitals) and transfers only models' updates to the main server. Predictive models created and trained on local nodes'/edges' data are participating in creating global/centralized federated models.

The central AI/ML component in a medical system is the main source of predictive models trained on a number of datasets from local edge nodes/hospitals, and the models are constantly updating when new data appears and getting more and more reliable and with higher prediction power.

## 4    Medical Decision Support Systems

In this section we will present two characteristic medical systems based on contemporary technological achievements.

### 4.1    Smart Ambient Intelligent Living Environments

Technological advancements and innovations can significantly support patients to efficiently cope with everyday activities [10].

Ambient Assisted Living (AAL) and Ambient intelligent (AmI) environments facilitate patients in their living space. They incorporate intelligent and flexible services to patients acting in their living space like: Sensors, Networks, Pervasive Ubiquitous Computing and AI, Unobtrusive Human-Computer Interfaces [20].
AAL and AmI encompass monitoring services that supports patients in their everyday activities and living habits, also suggesting them possible actions that can improve their QoL and wellbeing [3], [17]. Numerous sensors are located at different places, such as sensors for light control, home automation control, presence sensor, medication control, and others to collect patients' data and monitor their daily activities and behaviors (see **Error! Reference source not found.**).

To propose patients' personalized predictions, treatments, recommendations or even possibility of prevention some serious diseases a wide variety of data are collected from such smart environments and processed using advanced AI/ML methods.

The availability of the smart devices and wearable sensor technology [25] are prominent in a fast accumulation of patient's sensitive and complex health data.

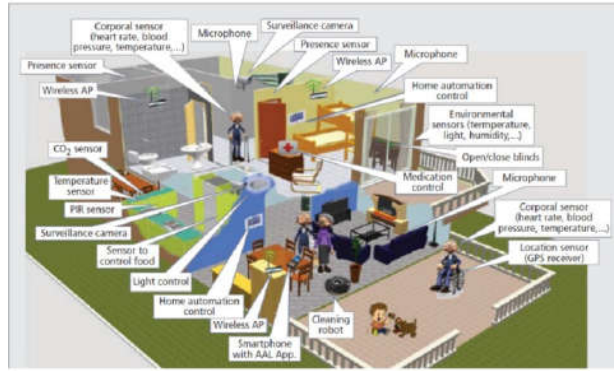

**Fig. 1.** AAL for elderly people's residence or house (from [17]).

### 4.2 Intelligent System for Supporting Cancer Patients

There are multiple challenges to be addressed by an AI system that aims to enhance clinical practice. A good AI engine with excellent analytical performance characteristics is not sufficient. Matters such as user experience, integration, security, privacy, etc. must also be addressed.

The EU-funded research and innovation project ASCAPE (https://ascape-project.eu/) presents an interesting proposition covering all aforementioned aspects of a system that aims to be in a position to enhance clinical practice.

Specifically, ASCAPE aims to provide doctors with an AI-powered tool that monitors and predicts the progression of QoL metrics corresponding to overall QoL and specific issues for a specific patient and offers suggestions for interventions that could improve outcomes. The ASCAPE Dashboard being the the primary interface was conceived as a tool that helps doctors better support cancer patients after their treatment by means of effective visual presentation of recorded and predicted data values and efficient and meaningful interaction. On the top of the ASCAPE personalized visualisations widget, the patient's overall QoL timeline is presented (higher is better) and on the bottom specific psychological, physiological and other QoL issues timelines (lower is better); just below the overall QoL or QoL issue timeline the various interventions (non-pharmacological & pharmacological) relevant at each point of time are also visualized (as straight-line segments).

The ASCAPE patient visualisations widget allows the doctor to get an overview of the patient QoL and the history of interventions without a litany of interactions. The default view provides both recorded data and predictions for the case that any currently active interventions remain so. Doctors can see how different choices of interventions affect the predictions for the patient's overall QoL and all QoL issues

ASCAPE, unlike the majority of similar clinically-targeted AI-focused research projects, paid particular attention to providing an easy pathway for integration with existing systems.

Another priority for ASCAPE is that hospitals on the one hand maintain control of their patient data and on the other are able to collaborate on building AI models capturing knowledge from multiple hospitals' patients. For this it relies on two different technologies: Federated ML (FL) and machine learning on homomorphic encrypted (HE) data.

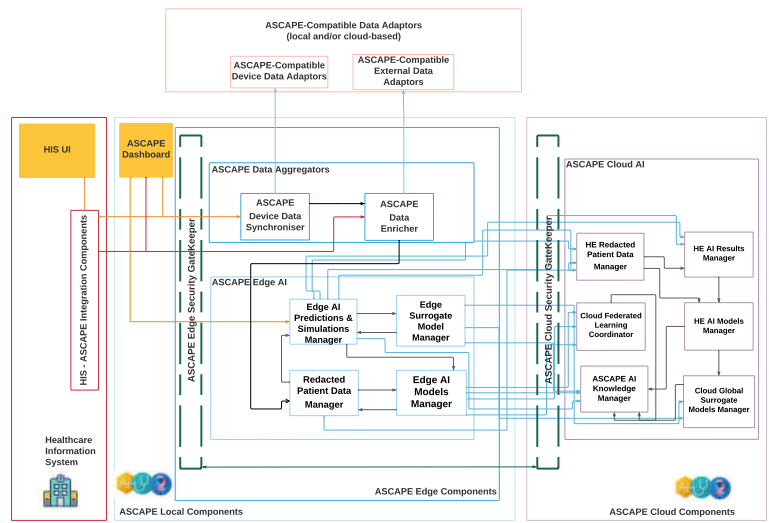

**Fig. 2.** The ASCAPE Framework Architecture.

**HIS-ASCAPE Integration Components.** These components allow an existing HIS to send its data (including both EHR and QoL questionnaire data) to ASCAPE (minimum integration requirement) and ideally also integrate the ASCAPE widgets and supporting backend code that provides the HIS with ASCAPE functionality identical to the stand-alone ASCAPE Dashboard's making the latter redundant and offering doctors the benefits of ASCAPE within the system they already use.

**The ASCAPE Dashboard.** A web application doctors may use, if ASCAPE is not sufficiently integrated into the HIS, in order to access ASCAPE functionality including AI-assisted monitoring of their patients' QoL status and recording information about proposed interventions, registering and de-registering a patient's wearable device.

**The ASCAPE Edge Components.** Installed locally at each hospital, these components collaborate with the HIS and the Dashboard on one end and, if so configured, with the ASCAPE Cloud which coordinates privacy-compliant collaborative model training with all participating hospitals and provides collaboratively training predictive models to all hospitals. More details of different edge node components are provided below. Note that all edge node components that interact with the ASCAPE cloud, any interactions are initiated from the edge node towards the cloud and never from the cloud to the edge node in order to fit as best as possible to firewall settings in place at hospitals IT environments behind which the edge node components are installed.

    **A. Edge Gatekeeper** - A component that provides TLS/SSL termination, access control and an additional layer of pseudonymization.

    **B. Data Aggregators** - Components that provide support to the task of sending to the ASCAPE Edge Node additional patient-related data not collected by the HIS but rather by ASCAPE-compatible Data Adaptors deployed locally or remotely.

    **C. Redacted Patient Data Manager** - The component responsible for storage of all patient data received from the HIS or the data aggregators within the Edge Node. It furthermore includes the extraction of patient specific inference requests and training datasets for the different target variables to train models. The component is also responsible for providing stored patient-related data to the HIS and/or the Dashboard.

    **D. Edge AI Models Manager** - The component responsible for training local models and for training global models with local data in collaboration with the ASCAPE Cloud, as well as for analytically evaluating models and choosing the ones that best fit local data. For each training dataset received from the Redacted Patient Data Manager several types of models are trained both on local data only as well in federated manner orchestrated by the Cloud Federated Learning Coordinator. For classification and regression tasks a range of ML algorithms is used. All locally trained models are stored in the component as well as any global model obtained from the ASCAPE cloud. The quality of the models is evaluated over the locally available datasets using adequate metrics.

    **E. Edge Surrogate Models Manager** - The component responsible for training local surrogate models (linear regression and decision trees) and for training global surrogate models for global predictive models with using the local data in collaboration with the ASCAPE Cloud; these surrogate models are trained to make the same predictions as the primary models (of the Edge AI Models Manager) but due to their nature (e.g. decision tree models) lend themselves to being used for explaining these predictions.

    **F. Edge AI Predictions & Simulations Manager** -The component that uses the locally available models (local models, global models from via federated learning) as well as the Homomorphic Encrypted (HE) models at the ASCAPE cloud to produce QoL-related predictions and intervention suggestions to the HIS and/or the Dashboard. The used models are those with the best evaluation over the local data and the predictions from the HE models are obtained by sending encrypted patient-specific inference requests to the ASCAPE cloud and decrypting locally the received encrypted prediction. Furthermore, the component is responsible to compute feature attributions in form of Shapley Values to allow to visualize the impact of the different features on the predicted target values.

    In addition to computing predictions and explanations, the component also pre-computes intervention suggestion: the goal is use the predictive capabilities of trained models and interventions of any kind for the patient and selected by the medical partners to provide for each patient with suggestions of interventions that have a positive effect on the predicted value. This is performed by simulations estimating the treatment effect of interventions (possibly also combinations of different interventions) and provide that

information for retrieval by the AASCAPE Dashboard to show it to the doctors treating the patients, which can then take a decision.

**The ASCAPE Cloud Components.** privacy-preserving machine learning technologies on the ASCAPE Cloud. This includes (i) the coordination and storage components for federated learning, (ii) the training, storage components for model training on homomorphically encrypted data and encrypted predictions, and (iii) the components used for collaborative surrogate model training specifically designed to compute linear regression and decision tree surrogate models.

**A. Cloud Gatekeeper** - A component that provides TLS/SSL termination and controls which Edge Nodes may collaborate with the ASCAPE Cloud. All communication requests initiated by any participating edge node passes through the gatekeeper to one of the functional components below.

**B. Cloud Federated Learning Coordinator** - This component coordinates the federated training of global predictive models based on the patient data available at each participating edge node. The same type of models as locally are trained in federated manner for classification and regression tasks. Typically, the edges in the federation are fixed, but this implies that hospitals cannot easily join a federation or leave it. Hence, specific training schemes were designed that allow for flexible addition and removal of edge node. The federated training is not initiated by the cloud federated learning coordinator, but rather by the edge nodes. If an edge needs a specific model and no global model is available in Cloud Knowledge Manager, it starts training locally and sends it as a first instance to the Cloud Federated Learning Coordinator. If a global model is available, the edge node updates it with its local training data and submits it again to the cloud. This mode of training is called incremental. If the Cloud Federated Learning Coordinator detects that more than one edge node want to train a model, it switches to semi-concurrent mode, where training happens in several rounds by collecting the trained or updated model from each edge node, creating an aggregated model by averaging and provide that model to all edge nodes for the next training round. All final trained global models are then forwarded to the Cloud Knowledge Manner.

**C. Cloud Knowledge Manager** - This component stores all available final global models on the cloud, from which they can be retrieved by the edge nodes. This way new edge nodes entering the federation can benefit from models previously trained on data from all other edge nodes. Furthermore, if an edge node decides to leave the federation, its contribution to the models is preserved thanks to the devised incremental and semi-concurrent training models.

**D. HE Redacted Patient Manager** - This component receives and stores the homomorphically encrypted training datasets from all edge nodes. The training datasets can be identified regarding cancer type and target variables and are combined to a single homomorphically encrypted dataset for each cancer type and target variable. These aggregated datasets are then forwarded to the HE AI Models Manager for training global homomorphically encrypted predictive models. The used homomorphic encryption scheme used in ASCAPE is a variant of the MORE (Matrix Operation for Randomization or Encryption) homomorphic encryption scheme, which relies on symmetric keys (a 2x2 matrix). The symmetric key is the same in all edge nodes where it is used to encrypt the patient data before uploading it the ASCAPE cloud.

**E. HE AI Models Manager** - This component stores all models trained on the aggregated homomorphically encrypted datasets. They can be retrieved by the HE AI Results Manager to provide encrypted predictions on encrypted inference requests submitted from the edge components.

**F. HE AI Results Manager** - The HE AI Results receives all encrypted inference requests for predictions from the different edge nodes. Based on the type, it retrieves the corresponding model from the HE AI Models Manager. If the model is not yet available, it waits until the model is available. The encrypted prediction is stored in the component in order to be retrieved by the edge node that submitted the request. The inference requests can be of different kinds: of course, any inference request in the edge node is also submitted to this component. However, during the computation of SHAPLEY values and the training of surrogate models further requests are created by the edge components and submitted to this component in order to determine these for the HE models.

**G. Cloud Global Surrogate Models Manager** - This component coordinates all activities to train global surrogate models, both for those obtained on plain text via federated learning as well as for global HE models. The training is initiated as soon as an edge node requests a surrogate model which is not yet trained. The Cloud Global Surrogate Models Manage then initiates the training both for linear regression and decision tree models. Meanwhile, the Edge Surrogate Model Manager creates the local training for the surrogate models by taking the local training dataset used for the global model, but labelling it using the predictions of the global model; in case this is an HE model, the labelling consists of submitting appropriate encrypted inference requests to the HE AI Results Manager and labelling the local dataset by decrypting the encrypted answer.

## 5      Conclusion

Growth of population and rapid technological development offer a variety of possibilities for implementing sophisticated and highly personalized medical services nowadays but in the future as well. Development of more and more powerful AI/ML algorithms, image processing, efficient big data processing, natural language processing, virtual and augmenter reality (VR/AR), IoT, agent technologies and other [11], offer a significant shift in medical and health domains [15].

All these possibilities direct medical research and practice in prominent directions [16]: more reliable and precise health analytics and predictive modeling [7], power data visualization techniques, tailored therapies, recommendations and interventions, personal user-friendly interfaces for communication [9] between different participants and stakeholders.

Ongoing and future research in the health domain needs extensive interdisciplinary and multidisciplinary collaborations. It is crucial to ensure good cooperation between software developers, data scientists and doctors, clinicians and healthcare providers to finally reach high quality final health systems/services.

Important aspect of future medical systems should take care of patients' cognitive and emotional behaviour and support adequate modelling in such systems. To empower friendly communication between system and patients and increase their motivation to follow medical recommendations the role of virtual e-coaches is extremely important. In this area agent technologies, holograms [12], AR/VR and metaverse definitely will play an essential role.

However, the near future is not so optimistic [19]. There are a lot of problems (like diverse, limited, and distributed patients' data sources, satisfactory but not fully reliable AI/ML models, rather slow big data processing mechanisms, integration of wide variety of multiple AI services) and personalized medicine has limitations that make it not revolutionary but evolutionary research domain.

**Acknowledgments.** This research was supported by the ASCAPE project. The ASCAPE project has received funding from the European Union's Horizon 2020 research and innovation programme under grant agreement No 875351.